\pdfoutput=1

\documentclass[a4paper,10pt]{article}

\usepackage[utf8]{inputenc}
\usepackage[T1]{fontenc}
\usepackage[frenchb,english]{babel}
\usepackage{textcomp}
\usepackage{amsmath}
\usepackage{amsfonts}
\usepackage{amssymb}
\usepackage{graphicx}
\usepackage{vmargin}
\usepackage[bookmarks=true,bookmarksnumbered=true,colorlinks=false,pdfborder={0 0 0}]{hyperref}
\usepackage{fancyhdr}
\usepackage{lastpage}
\usepackage{cases}
\usepackage[normal]{subfigure}
\usepackage{upgreek}
\usepackage{booktabs}

\setmarginsrb{25mm}{20mm}{25mm}{15mm}{12pt}{5mm}{0pt}{11mm}

\begin{document}


\pagestyle{fancy}
\chead{}
\lhead{\textit{B. Lenoir et al., 2013}}
\rhead{\textit{\thepage/\pageref{LastPage}}}
\cfoot{}
\renewcommand{\headrulewidth}{0pt}
\renewcommand{\footrulewidth}{0pt}



\title{Experimental demonstration of bias rejection from electrostatic accelerometer measurements}
\author{Benjamin Lenoir\textsuperscript{a}, Bruno Christophe\textsuperscript{a}, Serge Reynaud\textsuperscript{b} \\ \small \textsuperscript{a} \textit{Onera -- The French Aerospace Lab, 29 avenue de la Division Leclerc, F-92322 Ch\^atillon, France} \\ \small \textsuperscript{b} \textit{Laboratoire Kastler Brossel (LKB), ENS, UPMC, CNRS, Campus Jussieu, F-75252 Paris Cedex 05, France} \\ \\ \small Published in \textit{Measurement} 46:4 (2013) 1411-1420 \\ \small doi: \href{http://dx.doi.org/10.1016/j.measurement.2012.12.004}{10.1016/j.measurement.2012.12.004}}
\date{12 March 2013}
\maketitle



\begin{abstract}
In order to test gravitation in the Solar System, it is necessary to improve the orbit restitution of interplanetary spacecrafts. The addition of an accelerometer on board is a major step toward this goal because this instrument measures the non-gravitational acceleration of the spacecraft. It must be able to perform measurements at low frequencies with no bias to provide an additional observable of interest.

Since electrostatic accelerometers suffer a bias, a technological upgrade has been proposed by Onera. It consists in adding to an electrostatic accelerometer a rotating platform which allows modulating the signal of interest and retrieving it without bias after post-processing. Using this principle, a measurement method and a post-processing method have been developed. The objective of this article is to validate these methods experimentally. To do so, a horizontally controlled pendulum was used to apply a known signal to an accelerometer mounted on a rotating platform. The processing of the experimental data demonstrates the ability to make acceleration measurements with no bias. In addition, the experimental precision on the unbiased acceleration obtained after post-processing corresponds to the precision predicted theoretically.

\end{abstract}



\section{Introduction}\label{section:introduction}

The Roadmap for Fundamental Physics in Space issued by the European Space Agency (ESA) in 2010~\cite{esa2010roadmap} put an emphasis on gravitation tests in the Solar System with missions to the outer planets. In this framework, it recommends the development of accelerometers compatible with spacecraft tracking at the 10 pm.s$^{-2}$ level at low frequencies.

Indeed, interplanetary probes can be used as test masses whose trajectories are to be compared to theoretical predictions. Such a test has been performed by NASA with the Pioneer 10 and 11 probes: the outcome was a discrepancy with respect to the predictions of General Relativity \cite{anderson1998indication,anderson2002study,levy2009pioneer}.
But these two probes lacked an instrument which could disentangle gravitational effects from non-gravitational ones. In order to improve the experiment made by the Pioneer probes, several space missions have been proposed \cite{anderson2002mission,christophe2009odyssey,bertolami2007mission,dittus2005mission,johann2008exploring,wolf2009quantum}, many of them embarking an accelerometer. The OSS mission \cite{christophe2011oss} relies, for its fundamental physics objectives, on the Gravity Advances Package \cite{lenoir2011electrostatic}.
It is an instrument composed of an electrostatic accelerometer called MicroSTAR, based on Onera expertise in the field of accelerometry and gravimetry with CHAMP, GRACE, and GOCE missions~\cite{touboul1999accelerometers}, and a rotating platform, called Bias Rejection System~(BRS). In orbit technology is used with technological upgrades aiming at reducing power consumption, size and mass.

This instrument aims at measuring the non-gravitational acceleration of the spacecraft with a precision compatible with ESA requirement and with no bias. Thus, it provides an additional observable and allows removing, during the orbit restitution process, the effect of the non-gravitational forces on the trajectory, enhancing deep space gravitation tests as well as gravity field recovery~\cite{lenoir2010odyssey2IACproceedings}.
The measurements made by MicroSTAR or any other electrostatic accelerometer always include an intrinsic bias which may vary with time or temperature. It calls for an additional subsystem able to remove the bias through the modulation of the signal of interest. This is the role of the Bias Rejection System. In order to remove properly the bias of MicroSTAR, the Bias Rejection System rotates the accelerometer following a carefully designed periodical pattern~\cite{lenoir2011measuringSF2Aproceedings}.
In terms of measurement noise, this operation selects the noise of MicroSTAR around the modulation frequency. After post-processing and for a modulation period of $10$~min, it allows making absolute measurements with a white noise whose Power Spectrum Density (PSD) level is $10^{-10}$~m.s$^{-2}$.Hz$^{-1/2}$ with a cut-off frequency equal to $8.3\times10^{-4}$~Hz. This corresponds, for an integration time of $3$~h, to a precision of $1$~pm.s$^{-2}$ and an exact measurement accuracy~\cite{lenoir2012unbiasedASR}.
When taking into account the integration of the instrument in the spacecraft, and in particular the alignment accuracy ($\leq$ $1$~mrad), the positioning accuracy and the spacecraft self-gravity, a global precision of 10~pm.s$^{-2}$ is expected.

The spacecraft of the OSS mission has been specifically designed to reach this level of precision. The three main drivers taken into account are to (i) provide the lowest and most axisymmetrical gravitational field as viewed from GAP, (ii) make coincide the dry mass center of gravity, the propellant center of gravity, the radiation pressure force line and the GAP, and (iii) ensure a stable and reliable alignment between the GAP and the high gain antenna to ensure consistency between radio science data and non-gravitational acceleration measurements.
This led to the architecture presented in~\cite{christophe2011oss}: the platform is built as a flat ring and the GAP is on a settable plate at the center, held by thermally stable struts. In addition, four Hydrazine tanks are distributed symmetrically around the GAP, with fluidic interconnections. It allows balancing the quantity of propellant and thus controlling the position of the center of mass of the spacecraft.

The goal of the research project presented in this article is to validate experimentally the data processing method developed in \cite{lenoir2012unbiasedASR}. First, it will be shown that it is experimentally possible to completely remove the bias from the measurements. Then, the precision on the unbiased acceleration obtained after post-processing will be compared to the predicted precision given the noise of the devices used in this experiment.
These results will demonstrate the possibility to make unbiased acceleration measurements in real conditions using in space technology. Therefore, they are a key milestone toward embarking the Gravity Advanced Package on an interplanetary mission.

Since the Gravity Advanced Package is not available yet, commercial devices were used to mimic it: the ``measurement device'' which replaces the Gravity Advanced Package and the overall experimental setup is described in section~\ref{section:materials}. In section~\ref{section:methods}, the experimental methodology as well as the data processing is presented. Finally, the results are discussed in section~\ref{section:results}.

\section{Experimental setup}\label{section:materials}

\begin{figure}[ht]
\begin{center}
  \subfigure[The attitude of the pendulum is controlled by actuators (circled in red). The DM1 (circled in green) is used in the control loop to monitor the attitude of the pendulum. This set up allows removing the seismic noise. The two Q-Flex accelerometers (A and B) mounted on a Newport rotating platform (circled in blue) are used to perform the measurements in this experiment. The electronics are not visible on this picture.]{\includegraphics[width=0.45\linewidth]{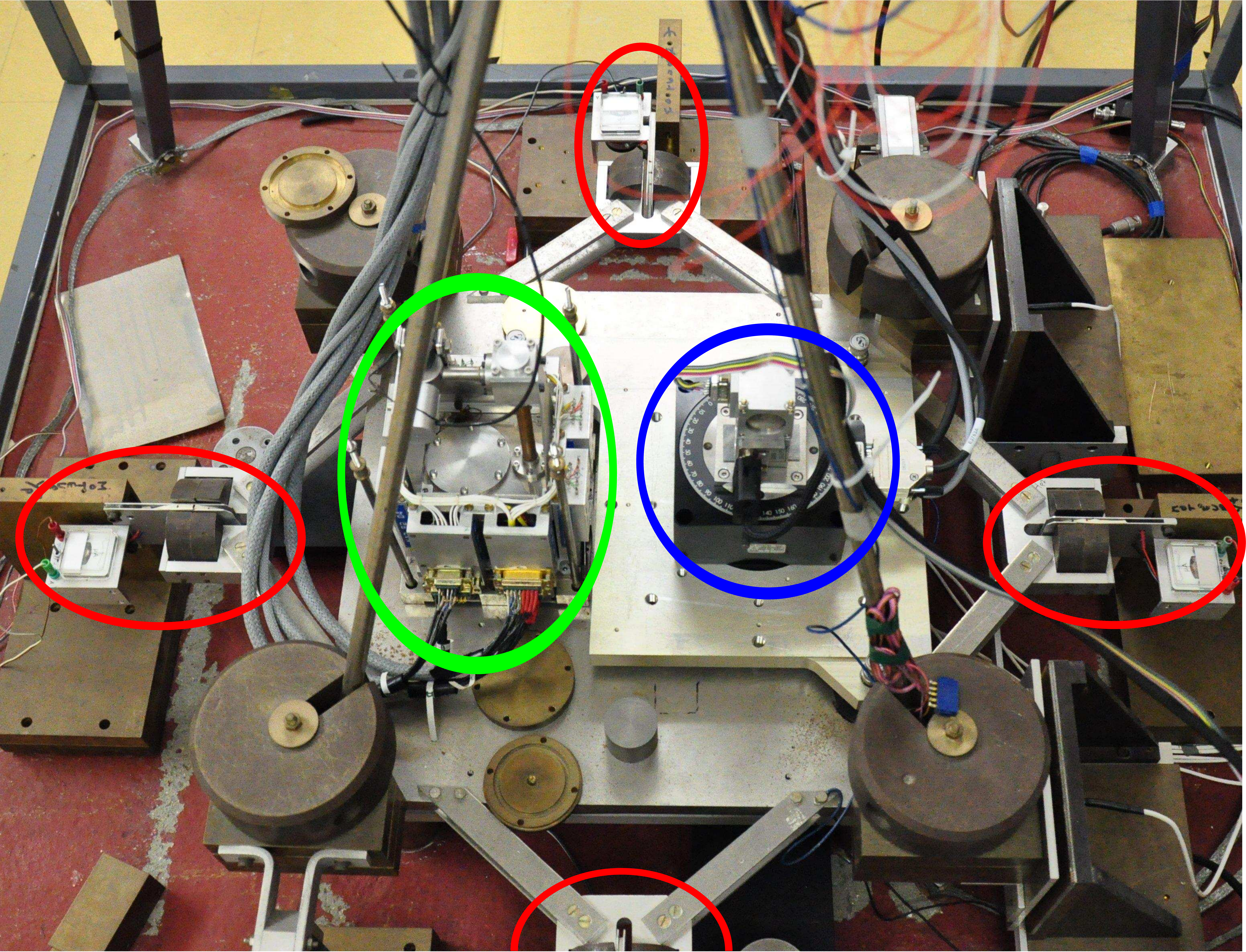}}
  \hskip 20pt
  \subfigure[The local reference frame ($X_0$,$Y_0$,$Z_0$) is defined such that $X_0$ is vertical. The angles $\phi$ and $\alpha$ (cf. Section~\ref{subsection:pendulum}) measure the inclination of the pendulum with respect to the horizontal plane.]{\includegraphics[width=0.45\linewidth]{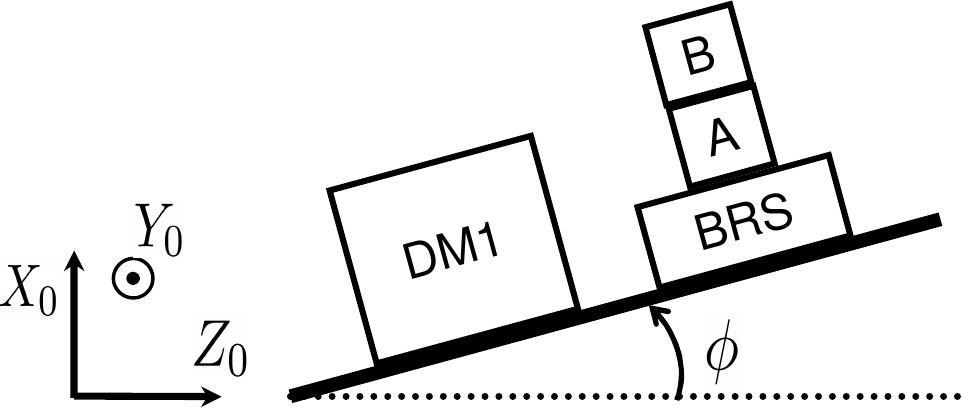}}
  \caption{Experimental setup}
  \label{fig:ExperimentalSetup}
\end{center}
\end{figure}

The overall experimental setup is shown in fig.~\ref{fig:ExperimentalSetup} shows. The ``measurement device'',  used to mimic the Gravity Advanced Package, is described in Section~\ref{subsection:accelerometer}. A pendulum controlled by an electrostatic accelerometer, called DM1, is used to impose a monitored external acceleration to the measurement device (cf. Section~\ref{subsection:pendulum}).

\subsection{Measurement device}\label{subsection:accelerometer}

\begin{figure}[ht]
\begin{center}
  \includegraphics[width=0.45\linewidth]{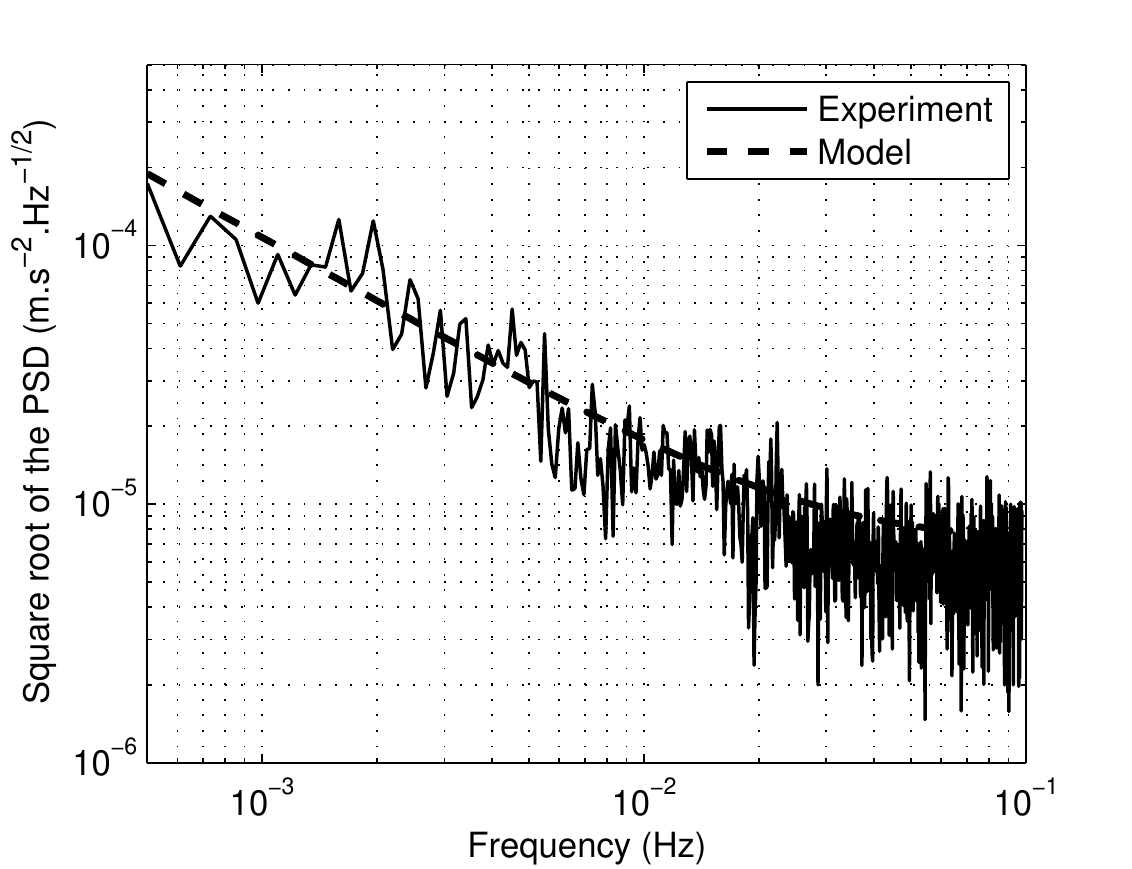}
  \caption{Square-root of the Power Spectrum Density (PSD) of the noise of the Q-Flex A and its measurement chain. The analytical model is given in equation \eqref{eq:psdQFlex}.}
  \label{fig:PSD_QFlex}
\end{center}
\end{figure}

The measurement device is composed of:
\begin{itemize}
 \item A rotating platform RGV100BL from Newport \cite{newport2011rgv100bl} which acts as the Bias Rejection System. The axis of rotation is called $x$ and is vertical. The rotation is parameterized by an angle called $\theta$. The angle repeatability is equal to $0.0003$°. Given the noise of the Q-Flex (cf. Fig.~\ref{fig:PSD_QFlex}), the rotating platform does not introduce uncertainty in the measurement process.
 \item Two Q-Flex QA-700 accelerometers from Honeywell \cite{honeywell2011qa700}. They are mono-axial accelerometers and are positioned orthogonally so as to have bi-axial measurements. The Q-Flex A and B measure local gravity along, respectively, the $z$ and $y$ axis of the frame attached to the rotating part of the rotating platform.
\end{itemize}

\begin{table}[ht]
 \renewcommand{\arraystretch}{1.3}
 \begin{center}
  \begin{tabular}{ l l }
   \toprule
   Q-Flex + converter & 0.8376 V.s$^2$.m$^{-1}$ \\
   gain & 1010.9 \\
   \hline
   Total & $\tilde{G}_A = $ 846.73 V.s$^2$.m$^{-1}$ \\
   \bottomrule
  \end{tabular}
 \end{center}
 \caption{Measured scale factor of the Q-Flex A}
 \label{table:QFlexA}
\end{table}

\begin{table}[ht]
 \renewcommand{\arraystretch}{1.3}
 \begin{center}
  \begin{tabular}{ l l }
   \toprule
   Q-Flex + converter & 0.8120 V.s$^2$.m$^{-1}$ \\
   gain & 990.54 \\
   \hline
   Total & $\tilde{G}_B = $ 804.32 V.s$^2$.m$^{-1}$ \\
   \bottomrule
  \end{tabular}
 \end{center}
 \caption{Measured scale factor of the Q-Flex B}
 \label{table:QFlexB}
\end{table}

\begin{figure}[ht]
\begin{center}
  \includegraphics[width=0.90\linewidth]{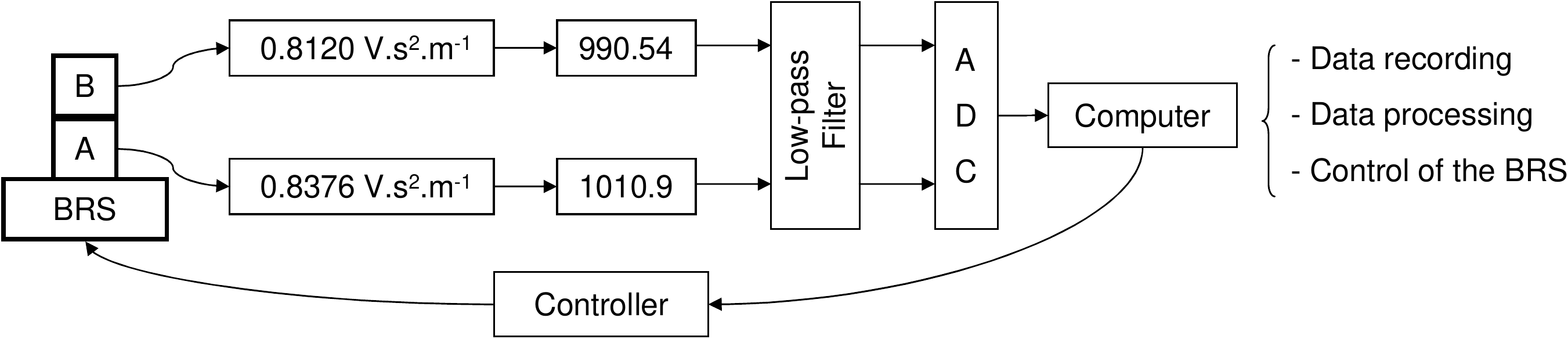}
  \caption{Schematic of the measurement chain. The computer is used for data recording and processing and for controlling the rotating platform. A low-pass filter is used before the Analog-to-Digital Converter (ADC) to avoid aliasing.}
  \label{fig:experimental_setup}
\end{center}
\end{figure}

As the output of the Q-Flex is a current proportional to the measured acceleration, the measurement chain of one Q-Flex accelerometer is composed of a current-voltage converter and a gain. In order to obtain the scale factor of the whole instrument, the calibration process is made in two steps. First, the scale factor of each accelerometer plus its current-voltage converter is computed. To do so, the measurement axis of the Q-Flex is set vertical so that the local gravity field $g$ is measured ; then it is flipped so as to measure $-g$. By subtracting and adding these two measurements and knowing the local gravity field, the gain and the bias of the Q-Flex with the current-voltage converter are known. Then, the scale factor of the gain is obtained by measuring the output (in V) knowing precisely the input (in V). The numerical values of the scale factors are summarized in tables \ref{table:QFlexA} and \ref{table:QFlexB}: $\tilde{G}_A$ and $\tilde{G}_B$ are the experimental values whereas $G_A$ and $G_B$ are
the exact ones, which are unknown in practice. Let introduce the measurement scale factors on each axis $k_y = (G_B-\tilde{G}_B)/\tilde{G}_B$ and $k_z = (G_A-\tilde{G}_A)/\tilde{G}_A$, which are small compared to 1.

The numerical values have been chosen such that the scale factors of the whole measurement chains for the Q-Flex are comparable to the ones of the DM1 (cf. next section). It is also possible to know approximately the numerical values of the bias of the whole measurement chains for the Q-Flex A and B, respectively $b_A = -224.50$ V and $b_B = -52.19$ V.

Finally, data need to be digitized for post-processing. To do so, the low-pass filter SR640 \cite{stanford2011programmable} is used to perform anti-aliasing before digitizing the data with the Analog-to-Digital Converter (ADC) NI 6033 card \cite{national2011ni}. The characterization of these devices in term of bias, scale factor and noise has been performed and is used in the data processing. For every experiment, the cut-off frequency is equal to the half of the sampling frequency in order to avoid aliasing \cite[p.~77]{papoulis1977signal}. The measurement chain is shown in fig.~\ref{fig:experimental_setup}.

In order to compare the predicted precision \cite{lenoir2012unbiasedASR} to the experimental one, it is necessary to characterize the noise of the measurement chain described above. Fig.~\ref{fig:PSD_QFlex} shows the measured Power Spectrum Density (PSD). It can be approximated by the analytical functions $S_{Q_A}(f) = S(f) / {\tilde{G^A}}^2$ and $S_{Q_B} = S(f) / {\tilde{G^B}}^2$ (in~m$^{2}$.s$^{-4}$.Hz$^{-1}$) corresponding respectively to the Q-Flex A and B with
\begin{equation}
 S(f) = 3.1925 \times 10^{-5} \ \mathrm{V}^2 . \mathrm{Hz}^{-1} + \frac{9.052 \times 10^{-8} \ \mathrm{V}^2 . \mathrm{Hz}^{0.6384}}{f^{1.6384}}
 \label{eq:psdQFlex}
\end{equation}

For frequencies above $2\times10^{-1}$ Hz, the analytical model does not fit well with the measured PSD. Indeed, at these frequencies, it is not the electronic noise of the Q-Flex which is predominant but the seismic noise filtered by the pendulum response function. It is however not a concern because these frequencies are too high to play a role in this experiment. For frequencies below $2\times10^{-1}$ Hz, the noise of the Q-Flex is far larger than the noise of the pendulum (cf. Fig.~\ref{fig:PSD_Pendulum}). This means that in this experiment, only the noise of the Q-Flex accelerometers plays a role.

\subsection{Horizontally controlled Pendulum}\label{subsection:pendulum}

\begin{figure}[ht]
\begin{center}
  \includegraphics[width=0.45\linewidth]{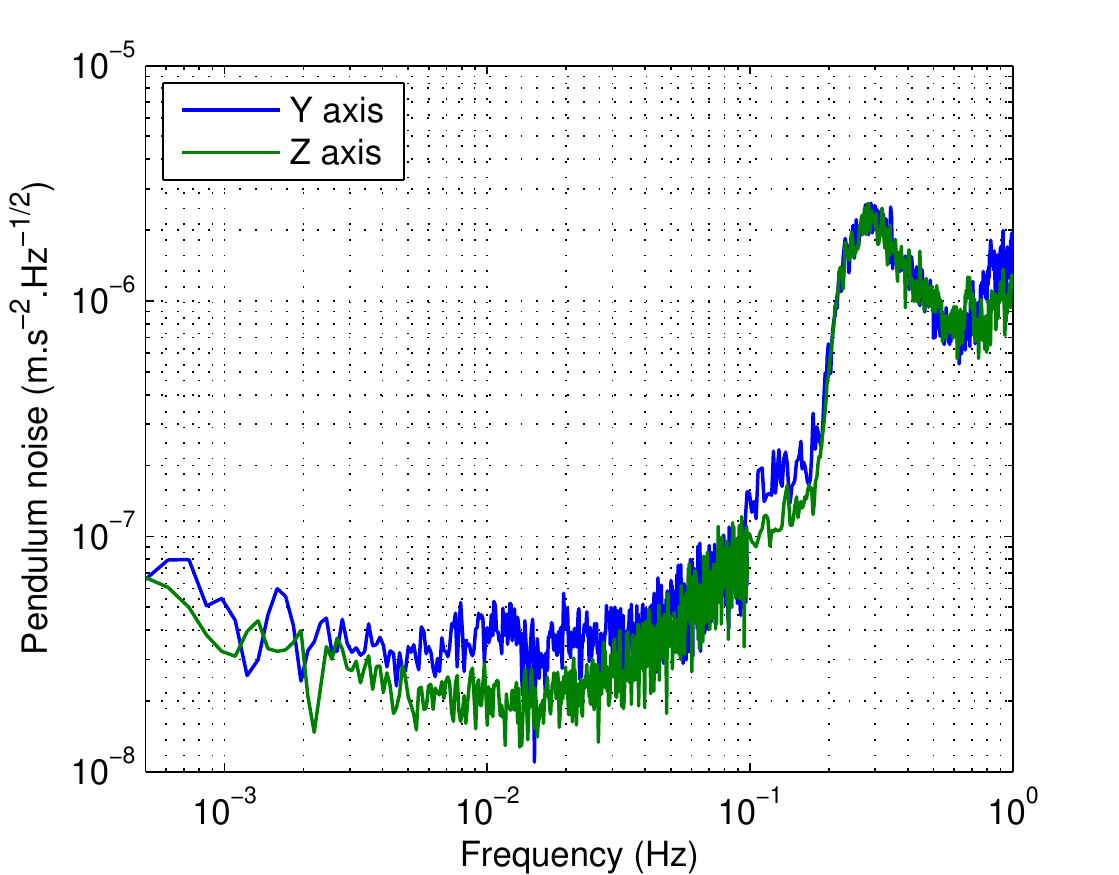}
  \caption{Square-root of the Power Spectrum Density of the pendulum noise. Below $2\times10^{-1}$ Hz, the pendulum is servo-looped on the outputs of the DM1 accelerometer and the noise level goes down to approximately $3\times 10^{-8}$ m.s$^{-2}$.Hz$^{-1/2}$. Above $2\times10^{-1}$ Hz, the pendulum is servo-looped on the ground and the seismic noise, filtered by the pendulum response function, is dominant.}
  \label{fig:PSD_Pendulum}
\end{center}
\end{figure}

The pendulum is used as a mean to deliver a known signal to the measurement device with a noise characterized by fig.~\ref{fig:PSD_Pendulum}. To achieve this level of noise, it is mounted on concrete decoupled from the ground. Moreover, the outputs of an electrostatic accelerometer, called DM1, are used as sensors: the servo-loop is designed such that the outputs in the horizontal plane ($Y$ and $Z$ axis of the DM1) are equal to zero. This accelerometer is an engineering model of the Aristoteles mission \cite{schuyer1992probing}. Table~\ref{table:DM1} gives the experimental values of the scale factors of the two horizontal axis of the DM1, $\tilde{G}_Y$ and $\tilde{G}_Z$. The exact values are named $G_Y$ and $G_Z$. As for the Q-Flex, let's also introduce the measurement scale factors on each axis $k_Y = (G_Y-\tilde{G}_Y)/\tilde{G}_Y$ and $k_Z = (G_Z-\tilde{G}_Z)/\tilde{G}_Z$.

\begin{table}[ht]
 \renewcommand{\arraystretch}{1.3}
 \begin{center}
  \begin{tabular}{ l l  }
   \toprule
   Y axis & $\tilde{G}_Y = 805$ V.s$^2$.m$^{-1}$ \\
   Z axis & $\tilde{G}_Z = 825$ V.s$^2$.m$^{-1}$ \\
   \bottomrule
  \end{tabular}
 \end{center}
 \caption{Scale factors of the two horizontal axis of the DM1 accelerometer \cite{chauvin1999star}.}
 \label{table:DM1}
\end{table}

Because of the bias $b_Y$ and $b_Z$ of the DM1 accelerometer along the $Y$ and $Z$ axis, the horizontal axis of the DM1 are perpendicular to local gravity with an offset. In addition, it is possible to incline the DM1 by a known angle. To do so, secondary entries on the $Y$ and $Z$ axis of the DM1, called $V_Y$ and $V_Z$ (in V), allow to put offsets. Given the scale factors of Table~\ref{table:DM1}, the inclination angle $\phi$ and $\alpha$ of the DM1, respectively around the axis $Y$ and $Z$, are equal to
\begin{subequations}
 \begin{numcases}{}
  \phi = \frac{b_Z-V_Z}{g G_Z} = \frac{1}{g} \left[\frac{b_Z}{G_Z} - \frac{1}{1+k_Z} c_Z \right]  \label{eq:defphi} \\
  \alpha = \frac{b_Y-V_Y}{g G_Y} = \frac{1}{g} \left[\frac{b_Y}{G_Y} - \frac{1}{1+k_Y} c_Y \right] \label{eq:defalpha}
 \end{numcases}
 \label{eq:defphialpha}
\end{subequations}
where $c_Y = V_Y/\tilde{G}_Y$ and $c_Z = V_Z/\tilde{G}_Z$ are offsets in term of acceleration. A typical amplitude for $V_Y$ and $V_Z$ is 1.5 V, which leads to an inclination of the order of $2\times 10^{-4}$ rad and a projection of local gravity in the $Y$-$Z$ plane of the order of $2\times 10^{-3}$ m.s$^{-2}$.

\section{Methods}\label{section:methods}

\subsection{Modeling of the experiment}\label{subsection:modeling}

As explained, the attitude of the DM1 with respect to the local reference frame is parameterized by the angles $\alpha$ and $\phi$. Because of the experimental setup, the alignment between the DM1 and the rotating platform is not perfectly known. Therefore, the attitude of the rotating platform with respect to the local reference frame is parameterized by the angles $\tilde{\alpha} = \alpha + \alpha_0$ and $\tilde{\phi} = \phi + \phi_0$, where $\alpha_0$ and $\phi_0$ are constant but unknown. For these experiments, $\alpha$ has been set to $0$, which means that only rotations of the pendulum around the axis $Y$ have been applied.

\begin{figure}[ht]
\begin{center}
  \includegraphics[width=0.45\linewidth]{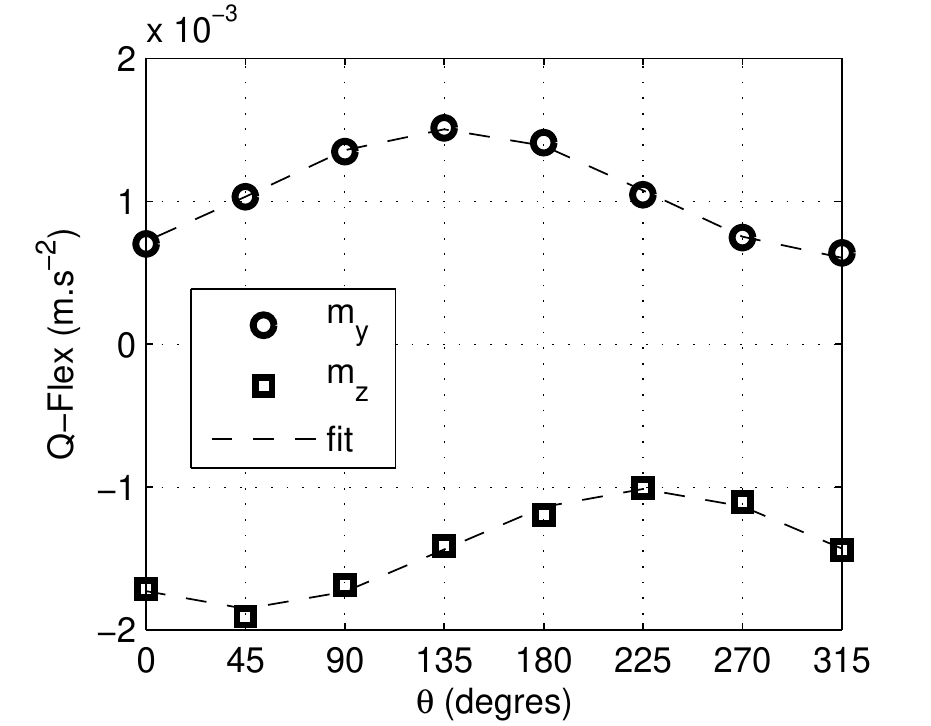}
  \caption{Outputs of the Q-Flex as a function of the angle $\theta$ for $\alpha = \phi = 0$. The amplitude of the curves is linked to the external acceleration and the offset of the curves is due to the bias and the wobble of the measurement device.}
  \label{fig:wobble}
\end{center}
\end{figure}

Calling $m_z$ and $m_y$ the measurements (in m.s$^{-2}$) made respectively by the Q-Flex A and B, they are equal at first order to
\begin{subequations}
 \begin{numcases}{}
  m_y = k_y \left[  a_Y \cos(\theta) + a_Z \sin(\theta) \right] + b_y \\
  m_z = k_z \left[ -a_Y \sin(\theta) + a_Z \cos(\theta) \right] + b_z
 \end{numcases}
 \label{eq:mesure_Q-Flex}
\end{subequations}
with $a_Y$ and $a_Z$ the external accelerations (in m.s$^{-2}$)
\begin{subequations}
 \begin{numcases}{}
  a_Y = g \alpha_0 \label{eq:defaY} \\
  a_Z = - g \tilde{\phi} \label{eq:defaZ}
 \end{numcases}
 \label{eq:defaext}
\end{subequations}
and $b_y$ and $b_z$ the measurement bias (in m.s$^{-2}$) of the whole ``measurement device'' (Q-Flex + rotating platform):
\begin{subequations}
 \begin{numcases}{}
  b_y = b_B/\tilde{G}_B + g \beta_B \\
  b_z = b_A/\tilde{G}_A - g \beta_A
 \end{numcases}
\end{subequations}
where $\beta_A$ and $\beta_B$ are the wobbles of the Q-Flex A and B on the rotating platform. The wobble is defined as the angle between the measurement axis of the Q-Flex and the plane perpendicular to the rotation axis of the rotating platform. The bias $b_A$ and $b_B$ are known experimentally (cf. Section~\ref{subsection:accelerometer}), but $\beta_A$ and $\beta_B$ are not, such that the values of $b_y$ and $b_z$ are not known a priori.

Equations \eqref{eq:mesure_Q-Flex} are identical to those studied in \cite{lenoir2012unbiasedASR}, which considered the measurement made by the Gravity Advanced Package (GAP) on an interplanetary spacecraft. Therefore, this experiment is representative of the measurements made by the GAP during a space mission and the conclusion drawn in this article can be applied to the space experiment, except that the noise of the Q-Flex is higher than the noise of the GAP. Fig.\ref{fig:wobble} shows the variation of $m_y$ and $m_z$ with the angle $\theta$ controlled by the rotating platform. The data are fitted, and it leads to the following numerical values:
\begin{subequations}
 \begin{numcases}{}
  m_y = \left[-2.94 \cos(\theta) -2.98 \sin(\theta) - 14.33 \right] \times 10^{-4} \ \mathrm{m.s}^{-2} \\
  m_z = \left[-3.34\cos(\theta) + 3.02 \sin(\theta) + 10.54 \right] \times 10^{-4} \ \mathrm{m.s}^{-2}
 \end{numcases}
 \label{eq:mesure_Q-Flex_num}
\end{subequations}
Using the offset of the two curves and knowing the bias of each Q-Flex, it is possible to determine the values of the wobbles : $\beta_A = -8.07 \times 10^{-5}$~rad and $\beta_B = 1.40 \times 10^{-4}$~rad.

\subsection{Description of the measurements}\label{subsection:measurement}

Concerning the applied external acceleration on the Z axis, $c_Z$, controlled by the inclination $\phi$ of the pendulum, two types of signals have been used. In one case, a constant inclination of the pendulum has been applied with different values for the offset $V_Z$ of the DM1 accelerometer: 0~V, 0.5~V, 1~V and 1.5~V\footnote{These offsets correspond to an acceleration, $c_Z$, in the $Y$-$Z$ plane of 0~m.s$^{-2}$, $6.06 \times 10^{-4}$~m.s$^{-2}$, $1.21 \times 10^{-3}$~m.s$^{-2}$ and $1.82 \times 10^{-3}$~m.s$^{-2}$ respectively.}. The data gathered with these signals are used in Section~\ref{subsection:uncertainty}. In the second case, the inclination of the pendulum had a sinusoidal and a triangular variation with an amplitude of 1 V and different periods: 0.75~min, 1~min, 5~min and 60~min. These data are use in Section~\ref{subsection:validation}.

\begin{figure}[ht]
\begin{center}
  \subfigure[]{\includegraphics[width=0.3\linewidth]{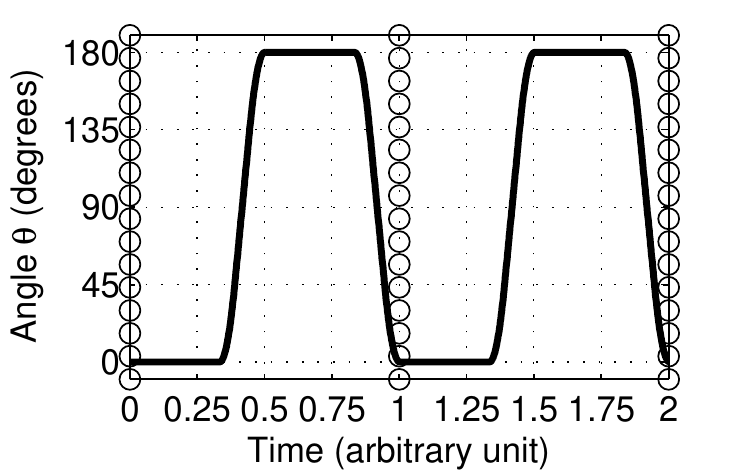}\label{fig:flip}}
  \subfigure[]{\includegraphics[width=0.3\linewidth]{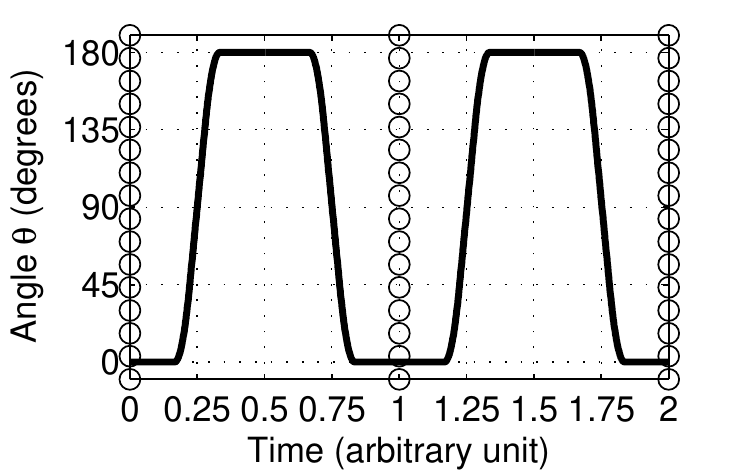}\label{fig:flipshift}}
  \subfigure[]{\includegraphics[width=0.3\linewidth]{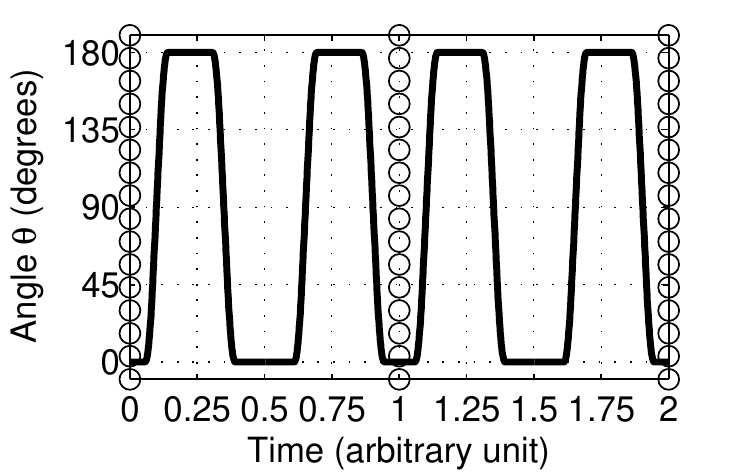}\label{fig:lin}}
  \caption{Calibration signals $\theta(t)$ used for the experiments. Two periods are represented, separated by circles ($\circ$). The total time spent rotating is always equal to $20$~seconds per period. The signal \ref{fig:flip} and \ref{fig:flipshift} are exactly similar except that the origin of the period is different, but they do not have the same properties as far as post-processing is concerned (cf. Section~\ref{subsection:validation}). The signal \ref{fig:lin} is different from the signal \ref{fig:flip} because it does not display a periodicity of 0.5~arbitrary unit.}
  \label{fig:calibration_signals}
\end{center}
\end{figure}

The time patterns used for the angle $\theta$, which parameterizes the rotation of the Q-Flex, are the one introduced in \cite{lenoir2012unbiasedASR}. They are represented in fig.~\ref{fig:calibration_signals}. Several modulation periods have been used: 1~min, 5~min, 20~min and 50~min. The characteristics of each signal in term of ability to remove the bias from the data is discussed in Section~\ref{subsection:validation}.

The sampling frequency used was mainly 10~Hz. Other sampling frequencies have been used so as to demonstrate that the precision on the post-processed quantities do not vary with this parameter under some conditions.

For all the data used in this article, the duration of the measurement was equal to 100 times the modulation period (e.g., for a modulation period of 50~min, the duration of the experiment was 3~days and 11~hours).

\subsection{Data processing}\label{subsection:processing}

\begin{equation}
  \Lambda_c = \mathrm{diag}[\cos(\theta_k)]
  \ \ \textnormal{and} \ \
  \Lambda_s = \mathrm{diag}[\sin(\theta_k)], \ k\in||1;N||.
\end{equation}
When the rotating platform is moving, it may induce additional signal due to off-centering as well as vibration. Therefore, only the measurement made when $\theta$ is constant are considered. Given the signals of fig.~\ref{fig:calibration_signals}, $\Lambda_s = 0$.

Because the scale factors $k_y$ and $k_z$ are unknowns, it is not possible to determine the quantities $a_Y$ and $a_Z$. Therefore, the quantities considered in the rest of this articles will be $\hat{a}_Y = (1+k_y) a_Y$ and $\hat{a}_Z = (1+k_z) a_Z$. Considering these quantities is natural because the values for the angle $\theta$ are 0 and 180 degrees (cf. Fig.~\ref{fig:calibration_signals}).

The overall goal of data processing is to find the projection of the vectors $\mathbf{\hat{a}_Y}$ and $\mathbf{\hat{a}_Z}$ on a vector subspace (of dimension $p_a \leq N$) whose basis is made of the column of a matrix $V_a \in \mathcal{M}_{N,p_a}(\mathbb{R})$, which are supposed to be orthogonal for the usual scalar product on $\mathbb{R}^N$. As a result, the goal is to find the numerical values of $({V_a}'V_a)^{-1} {V_a}' \mathbf{\hat{a}_Y}$ and $({V_a}'V_a)^{-1} {V_a}' \mathbf{\hat{a}_Z}$ knowing $\mathbf{m_y}$ and $\mathbf{m_z}$ ($M'$ is the matrix transpose of $M$).
In this article, the choice of $V_a$ (cf. Section~\ref{subsection:validation}) will allow retrieving the mean value of the acceleration without bias and/or the slope of the acceleration over one modulation period.

It has been demonstrated \cite{lenoir2012unbiasedASR} that it is possible to remove perfectly the bias from the measurements and that
\begin{subequations}
  \begin{numcases}{}
    {V_a}' \mathbf{\hat{a}_Y} = {V_a}' \Lambda_c \mathbf{m_y} \label{eq:a_simplified_Y}\\
    {V_a}' \mathbf{\hat{a}_Z} = {V_a}' \Lambda_c \mathbf{m_z} \label{eq:a_simplified_Z}
  \end{numcases}
  \label{eq:a_simplified}
\end{subequations}
under the following conditions
\begin{equation}
  {V_a}' \Lambda_c \mathbf{b_\kappa} = 0, \ \mathrm{with} \ \kappa\in\{\mathbf{y};\mathbf{z}\}.
  \label{eq:condition_demodulation_a}
\end{equation}

In the same way, it is possible to retrieve the bias of the instrument. As for the acceleration, the goal is to find the projection of the vectors $\mathbf{b_y}$ and $\mathbf{b_z}$ on a vector subspace (of dimension $p_b \leq N$) whose basis is made of the column of a matrix $V_b \in \mathcal{M}_{N,p_b}(\mathbb{R})$, which are suppose to be orthogonal for the usual scalar product on $\mathbb{R}^N$. Under the following conditions
\begin{equation}
  {V_b}' \Lambda_c \mathbf{a_\kappa} = 0, \ \mathrm{with} \ \kappa\in\{\mathbf{Y};\mathbf{Z}\}.
  \label{eq:condition_demodulation_b}
\end{equation}
it is possible to retrieve the bias:
\begin{subequations}
  \begin{numcases}{}
    {V_b}' \mathbf{b_y} = {V_b}' \mathbf{m_y} \\
    {V_b}' \mathbf{b_z} = {V_b}' \mathbf{m_y}
  \end{numcases}
  \label{eq:b_simplified}
\end{subequations}
As for the external acceleration, $V_b$ will be chosen such that the mean value of the bias and/or the slope of the bias over one modulation period are retrieved after post-processing.

It is a priori not possible to know if conditions \eqref{eq:condition_demodulation_a} and \eqref{eq:condition_demodulation_b} are fulfilled since the temporal evolution of the bias and of the external acceleration may not be controlled. In this experiment however, it is possible to assume that the bias belongs to the subspace generated by the columns of a matrix $\tilde{V}_b$ and that the external acceleration belongs to the subspace generated by the columns of a matrix $\tilde{V}_a$. Under these assumptions and given that $\Lambda_s=0$, the previous conditions \eqref{eq:condition_demodulation_a} and \eqref{eq:condition_demodulation_b} come down to ${V_a}'\Lambda_c\tilde{V}_b = 0$ and ${V_b}'\Lambda_c\tilde{V}_a = 0$.

Finally, it is essential to notice that the post-processed acceleration and bias are obtained separately and under different conditions. For example, it is possible to retrieve correctly the external acceleration without bias if ${V_a}'\Lambda_c\tilde{V}_b = 0$ whereas the post-processed bias will be mixed with the external signal because ${V_b}'\Lambda_c\tilde{V}_a \neq 0$.

\section{Results and discussion}\label{section:results}

Three goals are pursued with the experimental setup described above. First, a validation of the modeling made in Section~\ref{subsection:modeling} is presented. Second, it will be shown in Section~\ref{subsection:validation} that obtaining unbiased measurement is possible applying the post-processing scheme described above. Finally, in Section~\ref{subsection:uncertainty}, the experimental uncertainty of the unbiased measurements will be shown to follow the theoretical prediction.

\subsection{Calibration of the relative orientation}\label{subsection:calibration}

It is interesting to aggregate all the experimental data in order to calibrate the unknown scale factors. For each condition with a constant inclination of the pendulum $c_Z$, an experimental value for $\hat{a}_Z$ has been computed from the data by applying equation \eqref{eq:a_simplified_Z}. Considering equations \eqref{eq:defphi} and \eqref{eq:defaZ}, the theoretical link between these two quantities is
\begin{equation}
 \hat{a}_z = \frac{1+k_z}{1+k_Z} c_Z - (1+k_z) \left[ \frac{b_Z}{G_Z} + g \phi_0 \right].
 \label{eq:fit}
\end{equation}
Experimentally, the following numerical values can be computed from the data
\begin{equation}
 \hat{a}_z = 0.964 \times c_Z + 2.8757 \times 10^{-4} \ \mathrm{m.s}^{-2}.
\end{equation}
The experimental value of $(1+k_z)/(1+k_Z)$, which is close to 1, demonstrates that the numerical values $\tilde{G}_A$ and $\tilde{G}_Z$ used to process the data are close to the real ones. Unfortunately, this fit does not allow to compute independently the value of $b_Z$ and $\phi_0$.

\subsection{Experimental validation of the demodulation process}\label{subsection:validation}

In Section~\ref{subsection:processing}, the equations \eqref{eq:condition_demodulation_a} and \eqref{eq:condition_demodulation_b} gave the conditions under which it is possible to retrieve the unbiased acceleration and the bias of the instrument from the data. To be more specific, let's introduce the following matrices:
\begin{equation}
   V_1 =
  \begin{bmatrix}
    \mathbf{1}_q &         & \mathbf{0}      \\
                 &  \ddots &                 \\
    \mathbf{0}   &         & \mathbf{1}_q \\
  \end{bmatrix}
\end{equation}
and
\begin{equation}
  V_2 =
  \begin{bmatrix}
    \mathbf{1}_q &         & \mathbf{0}   & \mathbf{t}_q &         & \mathbf{0}      \\
                 &  \ddots &              &                 &  \ddots &                 \\
    \mathbf{0}   &         & \mathbf{1}_q & \mathbf{0}      &         & \mathbf{t}_q \\
  \end{bmatrix}
\label{eq:def_V_2}
\end{equation}
where $\mathbf{1}_q$ is a matrix of $\mathcal{M}_{q,1}(\mathbb{R})$ whose coefficients are $1$, and $\mathbf{t}_q$ is a matrix of $\mathcal{M}_{q,1}(\mathbb{R})$ such that ${\mathbf{t}_q}_k = (k-q/2)\delta t$. $q$ is the number of data points during one period of the calibration signal. Assuming that a vector $\mathbf{x}$ belongs to the subspace generated by the matrix $V_2$ means that it is an affine function of time on each modulation period.
For a signal whose characteristic duration of variation is long compared to the period of the calibration signal, it is legitimate to assume that this signal can be approximated locally by an affine function and therefore that it belongs to the subspace generated by the matrix $V_2$ (it will be done below with a sinusoid).
\begin{itemize}
 \item For the calibration signal of fig.~\ref{fig:flip}: $V_1 \Lambda_c V_1 = 0$ and $V_1 \Lambda_c V_2 \neq 0$. It means, for example, that it is not possible to recover correctly the mean bias over a modulation period if the external acceleration is an affine function of time on each modulation period (cf. Fig.~\ref{fig:DemFlip1}).
 \item For the calibration signal of fig.~\ref{fig:flipshift}: $V_1 \Lambda_c V_2 = 0$ and $V_2 \Lambda_c V_2 \neq 0$. Fig.~\ref{fig:DemFlip2} provides an illustration of the case.
 \item For the calibration signal of fig.~\ref{fig:lin}: $V_2 \Lambda_c V_2 = 0$. In this case, it is possible to recover correctly the mean and the slope of the unbiased acceleration and of the bias over a modulation period under the assumption that they are affine functions of time on each period of the calibration signal.
\end{itemize}

\subsubsection{Perfect demodulation}\label{subsubsection:perfect}

\begin{figure}[!ht]
 \begin{center}
  \subfigure[Post-processed measurements giving the mean acceleration and the mean bias over a period of the calibration signal. Equation \eqref{eq:fit} was used to correct for the unknown alignment coefficients.]{\includegraphics[width=0.45\linewidth]{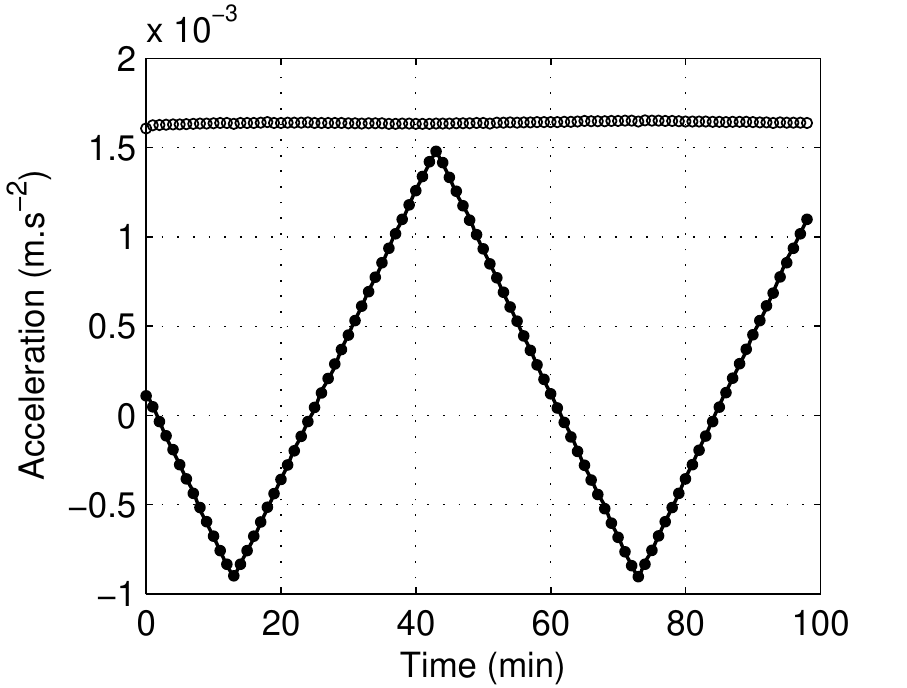}\label{fig:DemTriMean}}
  \hskip 20pt
  \subfigure[Difference between the post-processed mean acceleration and the real mean acceleration.]{\includegraphics[width=0.45\linewidth]{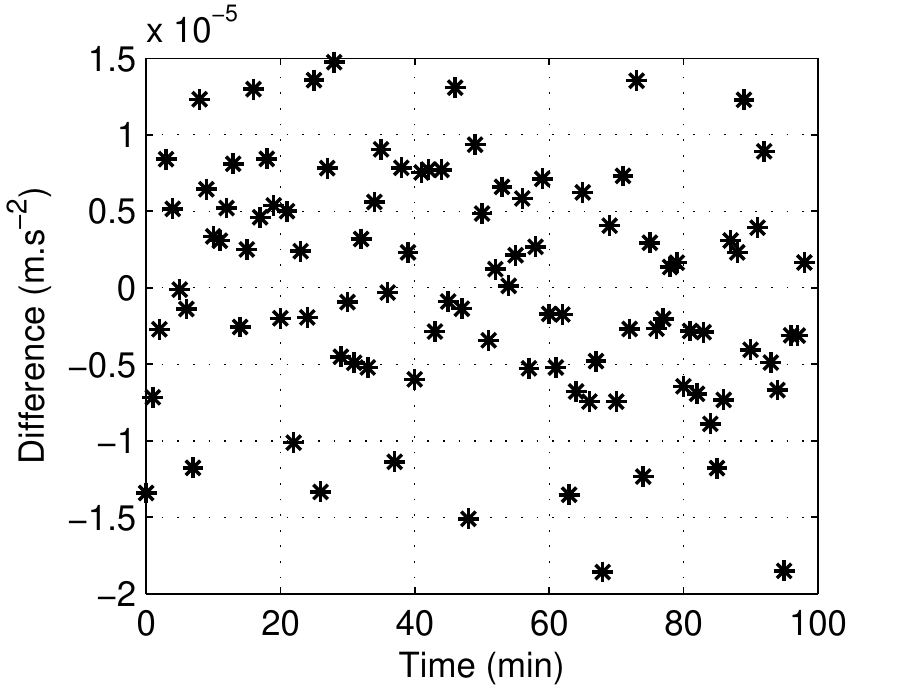}\label{fig:DemTriDiff}}
  \vskip 10pt
  \subfigure[Post-processed measurements giving the slope of the acceleration and of the bias over a period of the calibration signal]{\includegraphics[width=0.45\linewidth]{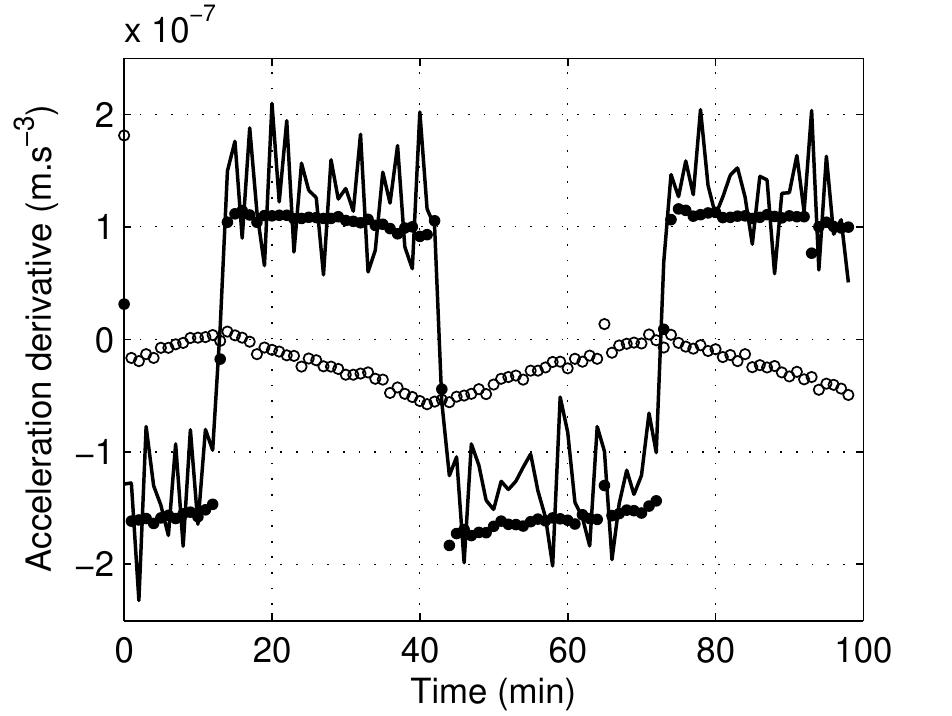}\label{fig:DemTriSlope}}
  \caption{Experimental conditions: the external signal is a triangular variation with a period of 60 min and an amplitude $V_z = 1$ V; the calibration signal is the signal of fig.~\ref{fig:lin} with a period of 1 min. In fig.~\subref{fig:DemTriMean} and \subref{fig:DemTriSlope}, the line (-) represent the known external signal, the dots ($\bullet$) represents the post-processed acceleration, and the circles (o) represented the post-processed bias.}
  \label{fig:DemTri}
 \end{center}
\end{figure}

First, let consider a case for which $V_2 \Lambda_c V_2 = 0$. To do so, the external signal is a triangular variation and the modulation signal is the one given in fig.~\ref{fig:lin}. The results of the data processing are plotted in fig.~\ref{fig:DemTri}.

Several features shows that post-processing allows recovering correctly the mean and the slope of the acceleration and bias over one period of the calibration signal. First the post-processed mean bias is constant, as expected since the temperature was constant (cf. fig.~\ref{fig:DemTriMean}). Second, the post-processed mean acceleration precisely corresponds to the external one with no bias, as shown by fig.~\ref{fig:DemTriDiff}. Finally, fig.~\ref{fig:DemTriSlope} shows that it is also possible to retrieve the slope of the acceleration.
In this figure however, it is possible to see that the slope of the bias has a variation which looks like the one of the external signal: it means that the conditions \eqref{eq:condition_demodulation_b} are not perfectly verified. This may come from the fact that according to fig.~\ref{fig:DemTriSlope}, the slope of the external acceleration is not perfectly constant but has random variations which interfere.

\subsubsection{Interference of the signal and the bias during post-processing}\label{subsubsectio:mix}

To illustrate the behavior of post-processing when the conditions \eqref{eq:condition_demodulation_a} and \eqref{eq:condition_demodulation_b} are not respected, let's consider the following experimental conditions: the external signal is a sinusoidal variation with a period of 60 min with a peak-to-peak value for $V_z$ of 1 V; the calibration signal is the signal of fig.~\ref{fig:flip} with a period of 5 min.

\begin{figure}[ht]
 \begin{center}
  \subfigure[Post-processed mean quantities over a period of the calibration signal. The data processing was made using the calibration signal of Fig.~\ref{fig:flip}. The peak-to-peak value of the variation of the post-processed bias is equal to $3.06 \times 10^{-4}$ m.s$^{-2}$.]{\includegraphics[width=0.45\linewidth]{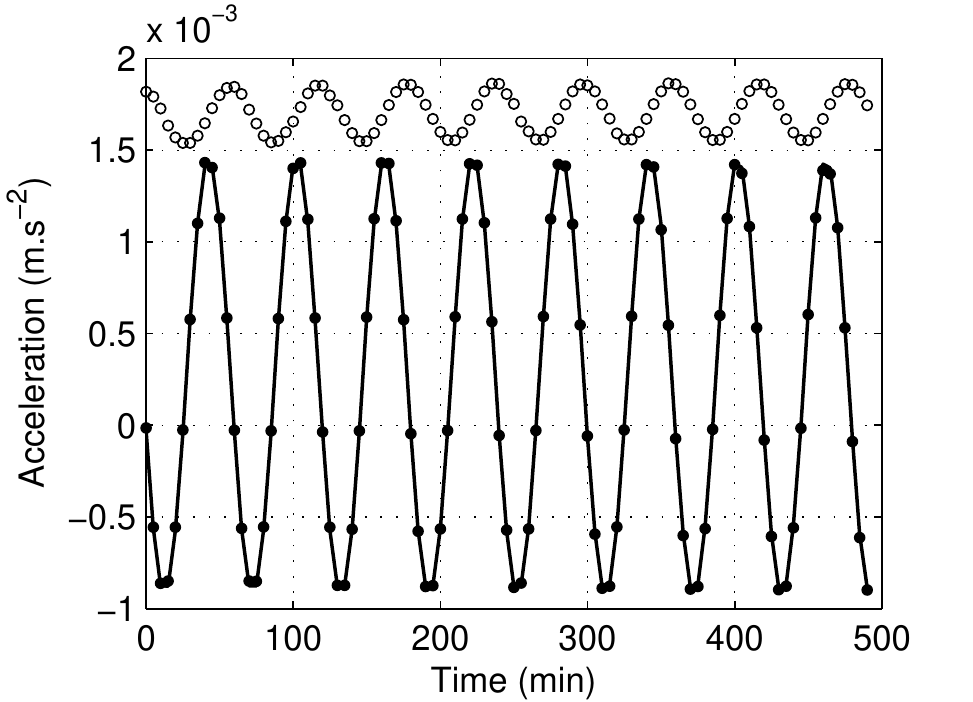}\label{fig:DemFlip1}}
  \hskip 20pt
  \subfigure[Post-processed mean quantities over a period of the calibration signal. The data processing was made using the calibration signal of Fig.~\ref{fig:flipshift}. The bias is not perfectly constant because the approximation that the sinusoid is affine with time on each modulation period is not perfectly correct, especially at the extrema.]{\includegraphics[width=0.45\linewidth]{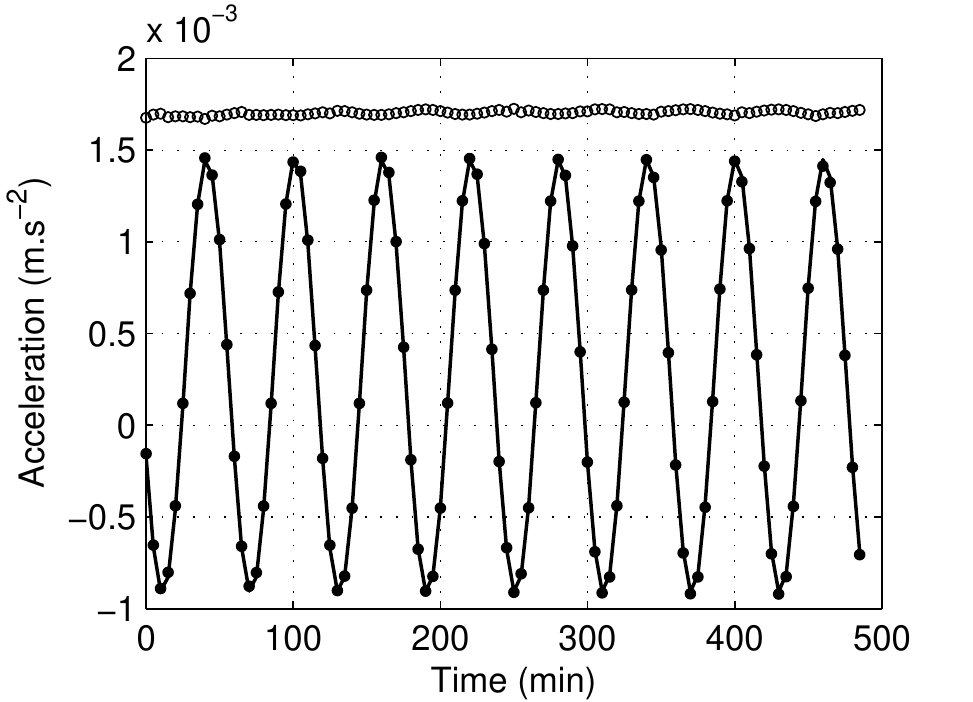}\label{fig:DemFlip2}}
  \caption{Experimental conditions: external signal is a sinusoidal variation with a period of 60 min with an amplitude $V_z = 1$ V; the calibration signal is the signal of Fig.~\ref{fig:flip} with a period of 5 min. The line (-) represents the known external signal, the dots ($\bullet$) represents the post-processed acceleration, and the circles (o) represented the post-processed bias. Equation \eqref{eq:fit} was used to correct for the unknown alignment coefficients.}
  \label{fig:DemFlip}
 \end{center}
\end{figure}

The data are processed in two different ways. In one case, the signal of fig.~\ref{fig:flip} is used, and in the other case, the period frame is shifted by 1/4 of a period and the data are processed using the signal of fig.~\ref{fig:flipshift}. The results are plotted in fig.~\ref{fig:DemFlip}.

Because the sinusoid cannot be considered as a constant function on each modulation period, $V_1 \Lambda_c \mathbf{a_\kappa} \neq 0$ for the signal \ref{fig:flip}. As a result, the post-processed mean bias is the addition of the real mean bias and an additional term, which is $({V_1}'V_1)^{-1} {V_1}'\Lambda_c \mathbf{\hat{a}_Z}$. This is a column vector and each value correspond to a sampling point. It can be approximated by the value of the following function computed at the sampling points:
\begin{equation}
 h(t) = \frac{1}{\tau} \left[ \int_{t}^{t + \tau/2} c_Z^0 \cos(\omega \lambda) d\lambda - \int_{t + \tau/2}^{t + \tau} c_Z^0 \cos(\omega \lambda) d\lambda \right] = \frac{2 c_Z^0}{\omega \tau} \left[1-\cos\left(\frac{\omega \tau}{2}\right)\right] \sin\left(\omega t + \frac{\omega \tau}{2}\right)
\end{equation}
where $\tau$ is the modulation period, $\omega$ is the pulsation of the sinusoid and $c_Z^0$ its amplitude. For the experimental conditions of fig.~\ref{fig:DemFlip1}, the peak-to-peak value of $h$ is $3.15 \times 10^{-4}$ m.s$^{-2}$ and its time shift is 17.5 min. It corresponds perfectly to the sinusoidal part of the post-processed bias of fig.~\ref{fig:DemFlip1}.

The result of the data processing with the calibration signal \ref{fig:flipshift} are plotted in fig.~\ref{fig:DemFlip2}. In this case, because $V_1 \Lambda_c \mathbf{a_\kappa} = 0$ for the the signal \ref{fig:flipshift}, there is no mix of the external acceleration with the bias: it is possible to retrieve the correct values of the mean bias and of the mean acceleration after post-processing.

\subsection{Uncertainty on the measurements: comparison to theoretical prediction}\label{subsection:uncertainty}

One of the important goal of this experiment is to be able to verify experimentally the precision on the mean unbiased acceleration. To do so, data acquired with a constant inclination of the pendulum are used and aggregated. The theoretical precision on the mean unbiased acceleration for an integration time $T$ measured after data processing can be expressed using the PSD of the Q-Flex A noise (cf. eq. \eqref{eq:psdQFlex}) and the calibration signal. In this section, the focus will be on the calibration signal \ref{fig:lin} because it is the most appropriate for our purpose, as demonstrated in the previous section.
The complete formula giving the uncertainty can be found in \cite{lenoir2012unbiasedASR} and is used for data processing. For simplicity however, let us recall the simplified expression giving the uncertainty $\sigma$ for an integration time $T$ and a modulation period $\tau$:
\begin{equation}
  \sigma(\tau,T) \approx \sqrt{\frac{1}{T} S_Q\left(\frac{1}{\tau}\right)}
  \label{eq:uncertainty}
\end{equation}
The general formula is an integral on the frequency of the noise PSD multiplied by the square of the norm of the Discrete Time Fourier Transform (DTFT) of the calibration signal. Moreover, since it has been shown \cite{lenoir2012unbiasedASR} that the noise on post-processed acceleration is a white noise, $\sigma(\tau,T_1) = \sqrt{T_2/T_1} \times \sigma(\tau,T_2)$. Therefore, the only integration time considered in the following is 1 hour.

\begin{figure}[ht]
\begin{center}
  \includegraphics[width=0.45\linewidth]{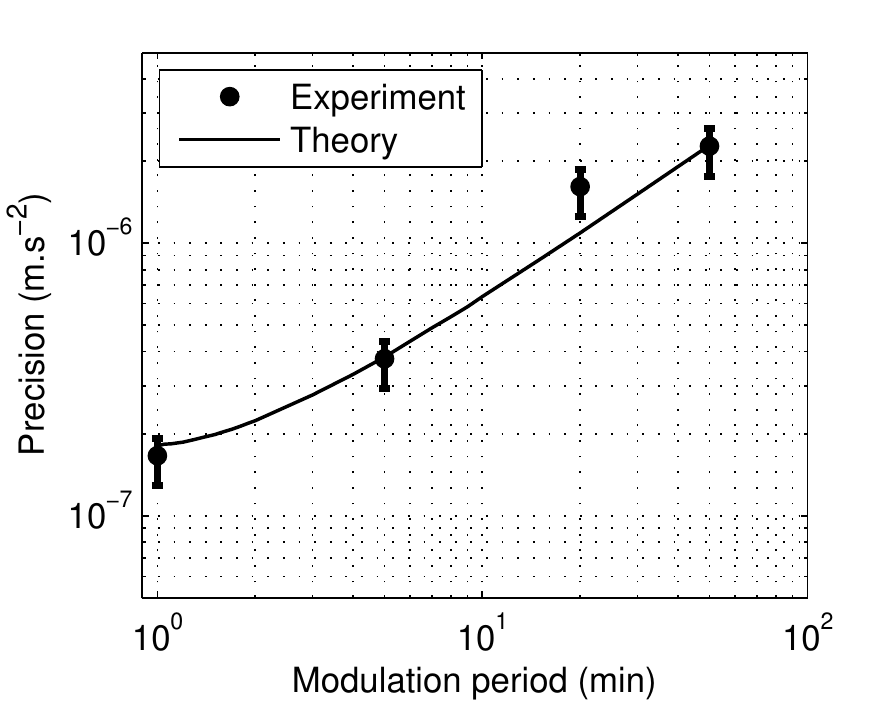}
  \caption{Precision ($1\sigma$) of the mean unbiased acceleration as a function of the period of the calibration signal for an integration time of one hour. The bars show the confidence interval at 99~\%. The data have been obtained with a sampling frequency of 10~Hz and with the calibration signal described in fig.~\ref{fig:lin}. Each point comes from the statistic over 200 to 700 data points depending on the modulation period. The theoretical uncertainty is also plotted using the general formula given in \cite{lenoir2012unbiasedASR}.}
  \label{fig:ExpUncertainty}
\end{center}
\end{figure}

This theoretical value for the uncertainty is compared to the experimental data in fig.~\ref{fig:ExpUncertainty}. The experimental data are obtained by computing the standard deviation of the experimental values of $\hat{a}_Z$ when the pendulum inclination is constant. Since the noise of the pendulum is much lower than the one of the Q-Flex, the measured values for $\hat{a}_Z$ changes from one modulation period to the other because of the measurement noise of the Q-Flex. By computing the standard deviation, one has access to the experimental precision on $\hat{a}_Z$.

Fig.~\ref{fig:ExpUncertainty} shows that the experimental values are in very good agreement with the theoretical prediction over a range of modulation period which is considered for the Gravity Advanced Package. Therefore, these experiments validate the theoretical approach developed in a previous article \cite{lenoir2012unbiasedASR} as well as the theoretical formula \eqref{eq:uncertainty}. The fact that the Q-Flex have a noise level much more higher than the noise level of MicroSTAR does not jeopardize this experimental validation of principle.

\begin{table}[ht]
 \renewcommand{\arraystretch}{1.3}
 \begin{center}
  \begin{tabular}{ l l }
   \toprule
   Frequency (Hz) & Precision (m.s$^{-2}$) \\
   \hline
   10  & $1.6631 \times 10^{-7}$ \\
   100  & $1.5718 \times 10^{-7}$ \\
   200  & $1.2524 \times 10^{-7}$ \\
   \bottomrule
  \end{tabular}
 \end{center}
 \caption{Influence of the sampling frequency over the precision of the unbiased acceleration. The precisions are given for an integration time of 1~hour. Experimental conditions: external signal is a constant with $V_Z = 0$~V; the calibration signal is the signal \ref{fig:lin} with a period of 1~min. The theoretical precision is $1.8073 \times 10^{-7}$~m.s$^{-2}$.}
 \label{table:frequency}
\end{table}

The effect of the sampling frequency has also been investigated. Because the harmonics of the calibration signal in the frequency domain decreases rapidly, it is expected that the precision does not depend on the sampling frequency as far as it is two order of magnitude larger than the modulation frequency. That is why the sampling frequency does not appear in the approximated formula \eqref{eq:uncertainty}. In these experiments, the larger modulation frequency is $1/60 = 0.0167$~Hz and the explored sampling frequencies are 10, 100 and 200~Hz, which are two orders of magnitude larger than 0.0167~Hz. The results are summarized in Table~\ref{table:frequency} and show that the precision does not depend significantly on the sampling frequency for the range explored in these experiments.

\section{Conclusion}\label{section:conclusion}

The set of experiments presented in this article was designed to validate the data processing scheme developed in order to remove the bias from the measurements made with an electrostatic accelerometer. To do so, two Q-Flex accelerometers mounted on a rotating platform were placed on a pendulum whose inclination was precisely controlled via a high-precision electrostatic accelerometer.

The data gathered provided an experimental validation of the method. First, the ability to recover, after post-processing, the mean and the slope of the bias and of the unbiased acceleration has been demonstrated with experimental data. Then, concerning the precision of the unbiased acceleration measurement, the experimental precisions have been compared to the theoretical predictions and a solid agreement has been obtained.

The experimental validation of this method devised to make unbiased acceleration measurement with an electrostatic accelerometer opens new opportunities. It will allow improved orbit restitution of interplanetary probes using an instrument with a strong heritage.

\section*{Acknowledgments}\label{section:acknowledgments}

We express our deep appreciation of the support from our colleagues at Onera: Damien Boulanger, Dominique Chauvin, Patrick Flinoise, Bernard Foulon, Yoann LeLay and Françoise Liorzou. We are also grateful to CNES (Centre National d'\'Etudes Spatiales) for its financial support.




\end{document}